\documentclass[preprint, amssymb, amsmath, amsfonts, superscriptaddress, aps, 30pt]{revtex4-2}
\usepackage{graphicx} 
\usepackage{dcolumn}
\usepackage{xcolor}
\usepackage{soul}
\usepackage{ulem}
\usepackage{bm}
\usepackage[utf8]{inputenc}    
\usepackage[T1]{fontenc}   
\usepackage{capt-of}
\usepackage{float}
\usepackage{etoolbox}  

\begin{document}
\title{Static Three-Dimensional Structures Determine Fast Dynamics Between Distal Loci Pairs in Interphase Chromosomes}

\author{Guang Shi}
\email{guang.shi.gs@gmail.com}
\author{Sucheol Shin}%
\affiliation{Department of Chemistry, The University of Texas at Austin, Austin, Texas 78712, USA}
\author{D. Thirumalai}
\email{dave.thirumalai@gmail.com}
\affiliation{Department of Chemistry, The University of Texas at Austin, Austin, Texas 78712, USA}
\affiliation{Department of Physics, The University of Texas at Austin, Austin, Texas 78712, USA}

\begin{abstract}
Live-cell imaging experiments have shown that the distal dynamics between enhancers and promoters are unexpectedly rapid and incompatible with standard polymer models. The discordance between the compact static chromatin organization and dynamics is a conundrum that violates the expected structure-function relationship.  We developed a theory to predict chromatin dynamics by accurately determining three-dimensional (3D) structures from static Hi-C contact maps or fixed-cell imaging data. Using the calculated 3D coordinates, the theory accurately forecasts experimentally observed two-point chromatin dynamics. It predicts rapid enhancer-promoter interactions and uncovers a scaling relationship between two-point relaxation time and genomic separation, closely matching recent measurements. The theory predicts that cohesin depletion accelerates single-locus diffusion while significantly slowing relaxation dynamics within topologically associating domains (TADs). Our results demonstrate that chromatin dynamics can be reliably inferred from static structural data, reinforcing the notion that 3D chromatin structure governs dynamic behavior. This general framework offers powerful tools for exploring chromatin dynamics across diverse biological contexts.
\end{abstract}

\maketitle

\newpage

\section{Introduction}
Over the last fifteen years, our understanding of chromatin organization has increased substantially, thanks to advances in experimental techniques, such as Chromosome Conformation Capture and its variants (collectively referred to as Hi-C) \cite{LiebermanAiden2009, Rao2014}, as well as multiplexed FISH and other fixed-cell imaging methods \cite{Faser15Microbiol,Bantignies2014,Beliveau2015natcomm,Wang2016science, Szabo2018scienceadvance,Bintu18Science, Su2020,Mirny22CSHL}. These studies, combined with computational modeling \cite{Barbieri2012pnas, Brackley2013pnas,Jost14NAR,Fudenberg2016, DiPierro2016, Shi2018, Liu2018ploscompbiol, Di21COGD, Thirumalai25ARB}, have revealed the organizational principles that underlie the three-dimensional structures of chromosomes at both the ensemble (obtained by averaging over a cell population) and the single-cell level.  For instance,  multiplexed-FISH experiments \cite{Finn2019, Bintu18Science, Su2020} and polymer theory~\cite{Shi19NatComm}  have been used to show that chromosomes exhibit extensive conformational heterogeneity at the single-cell level, reflecting the dynamical nature of their organization.  
The combination of experiments and polymer modeling has provided insights into the organization of interphase as well as mitotic chromosomes~\cite{Gibcus18Science,Dey23CellReports}.

Most experimental techniques rely on cell fixation methods, which are fundamentally limiting because they only probe static structures.  As a result,  our understanding of the potential structure-function relationship, which requires a quantitative understanding of the real-time dynamics of chromatin loci that control gene regulation (transcriptional bursting \cite{Fukaya2016cell, Bartman2016molcell}, for example) through enhancer (E)-promoter (P) communications \cite{Mach2023}, is limited. Recently, live-cell imaging experiments have probed the dynamics of chromatin. Such experiments fall into two categories: (i) Nucleosome positions are tracked without explicitly knowing their genomic identity. This can be used to measure dynamics at the multi-chromosomes and nucleus level \cite{Zidovska2013, Ashwin2019, Barth2020scienceadvance, Wagh2023scienceadvance}. (ii) Specific chromatin loci, limited to a small number, are marked and their movement as a function of time are tracked \cite{Chen2013cell}. This can be used to study the dynamics of specific genome regulatory elements, such as CTCF binding and enhancer-promoter interactions \cite{Chen2018naturegenetics, Alexander2019elife, Mach2022, Gabriele2022}.

In a recent notable development, Br{\"u}ckner et al. \cite{Bruckner23Science} employed a three-color labeling scheme to simultaneously probe the dynamics of several pairs of enhancers and promoters along with the transcription of the corresponding gene. The key results of the study, which investigated the one-point and two-point dynamical correlations of chromatin in \textit{Drosophila} cells, may be summarized as follows. (1) On the genomic scale, 58 kb $\le s \le $ 3.3 Mb where $s$ is the linear genomic length, chromosomes are compact,  resembling the fractal globule (FG) model \cite{Grosberg1988}. This implies that the mean distance, $r(s)$, between two loci should scale as $r(s) \sim s^{\nu}$ where $\nu = 1/3$ (fractal dimension is $1/\nu$). (2) However, the diffusion exponent, $\alpha$, characterizing the mean square displacement (MSD) of single chromatin loci and the two-point MSD are both approximately 0.5, which is close to the prediction using the Rouse model \cite{doi1988theory}. (3) Most notably, the relaxation time ($\tau$), associated with the two-point correlation, scales as a power law, $\tau \sim s^{\gamma}$ where $\gamma \approx 0.7$. Surprisingly, the measured ${\gamma}$ value is substantially smaller than the predictions based on both the Rouse model  ($\gamma=2$) and the Fractal Globule (FG) model ($\gamma=5/3$) \cite{Grosberg1988, LiebermanAiden2009, Halverson2014}. It is striking that the relaxation dynamics between pairs of loci occur on time scales that are substantially faster than predictions using dynamic scaling arguments that are based on the estimates of the mean separation between the loci using the FG or the Rouse model. The apparent lack of connection between the global static structures ($r(s)$ as a function of $s$) and the observed dynamic behavior requires a theoretical explanation.

Let us briefly explain the origin of the conundrum noted in the experiment \cite{Bruckner23Science}.  The typical relaxation time, $\tau_{r}$, of a polymer coil of length $s$ (measured along the polymer contour or genomic length) is given by $\tau_r = r^2(s) / D(s)$, where $r(s)$ is the characteristic size of the polymer, and $D(s)$ is the associated diffusion coefficient. If we assume that $D(s)$ obeys $D(s) \sim s^{-\theta}$ and $r^{2}(s) \sim s^{2\nu}$, we obtain the well-known relation $\tau_r \sim s^{2\nu + \theta}$ \cite{DeGennes1976}.  On the other hand, the time scale for single monomer diffusion at time $\tau_r$ must be consistent with $r(s)$, leading to the relation $\tau_r^{\alpha} \sim r^2(s) \sim s^{2\nu}$. Consequently, the diffusion exponent for a single monomer at intermediate timescales is $\alpha = 2\nu / (2\nu + \theta)$.  However, $\tau_r$ described above is difficult to quantify directly through experiments. Instead, Br{\"u}ckner et al. \cite{Bruckner23Science} measured the two-point dynamics using $M_2(t) = \langle||\boldsymbol{r}_{ij}(t)-\boldsymbol{r}_{ij}(0)||^2\rangle$ where $\boldsymbol{r}_{ij}(t)$ is the time-dependent  vector pointing from locus $i$ to locus $j$ and defined the relaxation time $\tau$ as the time at which $M_2(\tau)$ saturates at $\langle r^2(s)\rangle$. The scaling analysis shows that $\tau$ follows the same dependence on $s$ as $\tau_r$, namely, $\tau\sim s^{2\nu+ \theta}$. This shows that $\gamma=2\nu + \theta$. The scaling relation,

\begin{equation}
    \tau \sim s^{2\nu + \theta}\sim \langle r\rangle^{(2\nu+\theta)/\nu}
\label{EqScaling}
\end{equation}

\noindent links the relaxation time, $\tau$, between two loci with the mean spatial distance $\langle r\rangle$ or linear (genomic) distance $s$. For the Rouse chain, with $\theta=1$ and $\nu=1/2$, we find that $\tau \sim s^{2}$. For FG, with $\theta=1$ and $\nu=1/3$, it follows that $\tau\sim s^{5/3}$. The static structures  \cite{Bruckner23Science} suggests that $\nu=1/3$, consistent with the FG model. However, the experimentally measured exponent $\gamma \approx 0.7$ deviates from the expected value, $\gamma = 2\nu + \theta = 5/3$. Hence, there is a conundrum.

The failure of the Rouse or FG polymer models to account for the experimental observations~\cite{Bruckner23Science} prompted us to develop a new theory to explain the fast transcriptional dynamics (relaxation time between pairs of enhancer and promoter). Based on the discordance between global structure and relaxation dynamics one would be tempted to conclude that structure and dynamics are unrelated in chromatin. In this work, we first utilize our previous theory \cite{Shi2021PRX}  to calculate the three-dimensional (3D) structure of chromosomes using only the measured contact map. Using the ensemble of structures, we investigated the dynamics of distal pairs of chromatin loci.  Our model accurately predicts the experimental findings  using only the \textit{static} contact map as input, thus resolving the conundrum \cite{Bruckner23Science} by demonstrating that the chromatin dynamics can be derived from theory, provided the precise 3D structural ensemble is available. The unexpected scaling behavior observed in the experiment \cite{Bruckner23Science} arises from effective long-range interactions among chromatin loci, likely mediated by  transcription factors and cohesin. Because our theory is general, it is applicable to various cell types and species, enabling comparative investigations of chromosomal dynamics and mechanics in different species.

\section{Results}
\textbf{Outline of the Theory:} We developed a theory based on the supposition that knowledge of the static three-dimensional (3D) structure (namely, knowledge of all three-dimensional coordinates, $\{\boldsymbol{r}_i\}$ of the chromatin loci)  is sufficient to accurately predict the dynamics between arbitrary pairs of loci. The theory is executed in two steps. (i) We first use the measured (Hi-C or related methods) contact map to calculate the precise 3D structures \cite{Shi2021PRX} based on the maximum entropy principle, which yields the joint distribution function, $P^{\mathrm{MaxEnt}}(\{\boldsymbol{r}_i\})$.   The Hi-C contact map is used to calculate the mean distances ($\langle r_{ij} \rangle$) between loci $i$ and $j$ using polymer physics concepts~\cite{Bintu18Science,Shi18NatComm}.  The values of $\langle r_{ij} \rangle$ are needed to calculate  $P^{\mathrm{MaxEnt}}(\{\boldsymbol{r}_i\})$. The Lagrange multipliers (parameters), $k_{ij}$, in Eq. \ref{eq:eq2}, ensure that the mean distances between all pairs of loci match the calculated values using $P^{\mathrm{MaxEnt}}(\{\boldsymbol{r}_i\})$. (ii) By interpreting  $k_{ij}$ as spring constants in a harmonic potential in the chromatin network, we calculated the dynamical correlation functions using standard procedures used in the theory of polymer dynamics~\cite{doi1988theory}. The details follow.

\textbf{3D Structures from Hi-C Data:} The first step in the theory is the determination of the ensemble of 3D structures that are quantitatively consistent with the measured contact map. To this end, we used the polymer physics-based HIPPS (Hi-C-Polymer-Physics-Structures)  \cite{Shi2021PRX} and the related DIMES \cite{Shi2023NatComm} methods. The HIPPS  relates  the probability of contact, $\langle p_{ij} \rangle$, between loci $i$ and $j$, and the mean spatial distance $\langle r_{ij} \rangle$ separating them~\cite{Bintu18Science,Shi18NatComm} through the power law relation, $\langle  r_{ij} \rangle = \Lambda {\langle p_{ij} \rangle}^{-1/\alpha} $ with $\alpha \approx 4$. This relation, which was first reported in imaging experiments~\cite{Bintu18Science} and subsequently validated in simulations~\cite{Shi18NatComm}, differs from the predictions based on standard polymer models. To further validate this choice, we compared the inferred mean pairwise distances against imaging data over a range of $\alpha$ values. We find that $\alpha=4$ minimizes the root-mean-square deviation (fig. S11). With DIMES, we directly utilize the imaging data (coordinates of loci) to compute mean pairwise distances. In this work, we will refer our theory as HIPPS-DIMES. With $\langle  r_{ij} \rangle$ in hand, we formulate the maximum-entropy distribution as a function of the chromatin loci coordinates,

\begin{equation}\label{eq:eq2}
    P^{\mathrm{MaxEnt}}(\{\boldsymbol{r}_i\})\equiv P^{\mathrm{MaxEnt}}(\boldsymbol{r}_1, \boldsymbol{r}_2, \cdots)=\frac{1}{Z}\exp\bigg(-\sum_{i<j}k_{ij}||\boldsymbol{r}_i-\boldsymbol{r}_j||^2\bigg),
\end{equation}
\noindent where $Z$ is a normalization constant.  The elements,  $k_{ij}$, in Eq. \ref{eq:eq2} are the Lagrange multipliers which are determined to ensure that the average squared spatial distance between loci $i$ and $j$ matches the target values. We denote the matrix composed of all $k_{ij}$ elements as connectivity matrix, $\boldsymbol{K}$, with $K_{ij}=k_{ij}$ if $i\neq j$ and $K_{ii}=-\sum_{j\neq i}k_{ij}$. The central quantity of interest in our theory is the connectivity matrix, $\boldsymbol{K}$. To determine its elements, $k_{ij}$, we employed an iterative scaling algorithm designed to match the target $\langle r_{ij} \rangle$ values. The methodology is detailed in prior works \cite{Shi2021PRX, Shi2023NatComm}.

\textbf{Dynamics from $\boldsymbol{K}$:} Although the distribution $P^{\mathrm{MaxEnt}}(\boldsymbol{r}_1, \boldsymbol{r}_2, \cdots)$ (Eq. \ref{eq:eq2}) is calculated using the maximum-entropy principle, we interpret it as a Boltzmann distribution at unit temperature  ($k_BT$ is unity) with an effective energy, $H=\sum_{i<j}k_{ij}||\boldsymbol{r}_i-\boldsymbol{r}_j||^2$.  With this identification,  $k_{ij}$ may be interpreted as the spring constant between loci $i$ and $j$. Note that  $k_{ij}$ values are allowed to be negative, which indicates repulsion between chromatin loci. Despite the presence of negative $k_{ij}$ values, by construction $-\boldsymbol{K}$ remains positive semidefinite (although excessively large negative perturbations could violate semidefiniteness), thus ensuring that the probability distribution $P^{\mathrm{MaxEnt}}(\{\boldsymbol{r}_i\})$ is well-defined and normalizable. The interpretation that $P^{\mathrm{MaxEnt}}(\{\boldsymbol{r}_i\})$ resembles a Boltzmann distribution allows us to derive the inter-loci dynamics using the framework employed in the context of the Rouse model \cite{doi1988theory}. Therefore, the eigen-decomposition of the connectivity matrix $\boldsymbol{K}$  may be   used to calculate the normal modes. Each independent normal mode obeys the Ornstein-Uhlenbeck process. With this assumption,  dynamical quantities such as $M_2(t)$ can be expressed in terms of the eigenvalues and eigenvectors of $\boldsymbol{K}$ (see Supplementary Materials for details). It should be emphasized that the off‐diagonal “spring constants” associated with $\boldsymbol{K}$ should only be viewed as effective couplings to enforce the Hi-C or imaging-derived distance constraints. They are not literal molecular forces that act on large length scale.

To understand the loci relaxation dynamics, we define the two-point auto-correlation function $G_2(t)$ as $G_2(t)=\langle r^2(s)\rangle - M_2(t)/2$.  The dynamical scaling form of $G(t)$ should be $G_2(t)/G_2(0)\sim g(ts^{-b})$. At $t=\tau$, the curves collapse with $\tau s^{-b}\sim 1$ which leads to $b=2\nu+\theta$. It has been shown \cite{Polovnikov2018}  that for $s\ll N$, the scaling form is, $G_2(t)\sim t^{(2\nu-2)/(2\nu+1)}$. $G(t)$ also provides a well-defined way to define the relaxation time $\tau$  using $G(\tau)=1/e$. In our theory, the auto-correlation function for the pair of loci $i$ and $j$, $G_2^{ij}(t)$,  is given by,

\begin{equation}
\label{eq:eq3}
    G_{2}^{ij}(t)=\langle \boldsymbol{r}_{ij}(t)\boldsymbol{r}_{ij}(0)\rangle=3\sum_{p=1}^{N-1}(V_{pi}-V_{pj})^2 e^{-t/\tau_p}\big(-\frac{k_BT}{\lambda_p}\big)
\end{equation}

\noindent where $p$ is the normal mode index, and $\lambda_p$ and matrix $\boldsymbol{V}$ are the eigenvalues and eigenvectors of $\boldsymbol{K}$, respectively. The structure of Eq. \ref{eq:eq3} matches the Rouse model dynamics \cite{rouse1953theory, doi1988theory}, except in the chromatin $\boldsymbol{V}$ is non-trivial that requires numerical evaluation using the measured contact maps. The relaxation time for each normal mode is $\tau_p=-\xi/\lambda_p$ where $\xi$ is the friction coefficient, where $\xi$ represents an effective friction coefficient that incorporates the medium experienced by chromatin loci in the nucleus.  It is the only adjustable parameter in the theory and merely sets the overall time scale in all the dynamical predictions. The two-point MSD, $M_2(t)$, is calculated from $G_2(t)$ using $M_2(t)=2\langle r_{ij}^2\rangle-2G_2(t)$ where $\langle r_{ij}^2\rangle$ is the equilibrium mean squared spatial distance between the two chromatin loci. Note that evaluation of Eq. \ref{eq:eq3} requires only the properties of the matrix, $\boldsymbol{K}$. In the Rouse model, $\boldsymbol{K}$ is the polymer connectivity matrix, which is tri-diagonal. In the chromatin problem, it is calculated using Eq. \ref{eq:eq2} for which the experimental Hi-C/Micro-C contact map or imagined data is required.

\begin{figure}
\includegraphics[width=\columnwidth]{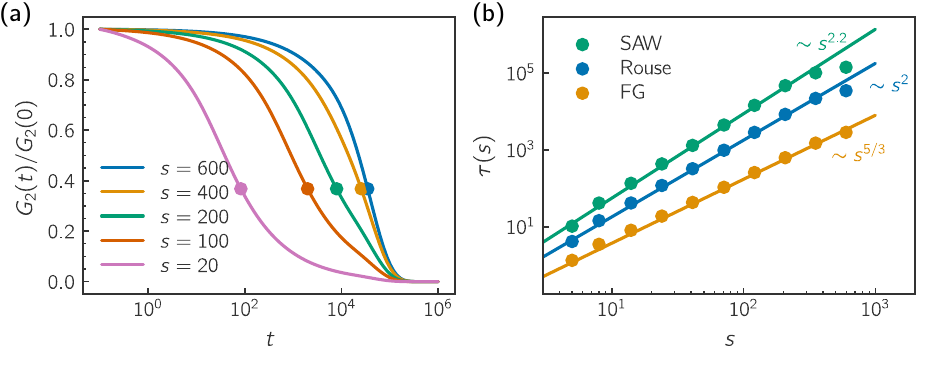}
\caption{\label{fig:fig2} \textbf{Two‐point relaxation dynamics in homopolymer models}. (a) Normalized two-point auto-correlation function, $G_2(t)/G_2(0)$, for pair of monomer of indices $i$ and $j$ with $s=|j-i|$. Loci pairs are selected symmetrically around the midpoint of the chain, $i=N/2-s/2$ and $j=N/2+s/2$. The results are for the Rouse model with chain length $N=1,000$. Solid circles mark the relaxation time $\tau$, defined as $G_2(\tau)/G_2(0)=1/e$. (b) Log-log plot of relaxation time $\tau$ as a function of the sub-chain length $s$ shows $\tau(s) \sim s^{x}$;  $x=2$ for the Rouse model, $x=5/3$ for the Fractal Globule (FG), and $x\approx 2.2$ for self-avoiding walk (SAW) chain. The chain length for all three models is $N=1,000$.}
\end{figure}

\textbf{Validating the theory:} To validate our theory, let us first show that HIPPS-DIMES correctly recovers the known scaling relations for the Rouse chain, self-avoiding walk (SAW), and FG. The mean spatial distance map for these models can be analytically constructed by using the well-known relations, $\langle r_{ij} \rangle =|i-j|^{1/2}$, $\langle r_{ij}\rangle =|i-j|^{3/5}$, and $\langle r_{ij}\rangle=|i-j|^{1/3}$, respectively. Using the analytic expressions for the mean distances, the first step in this theory is to determine $\boldsymbol{K}$ so that target mean pairwise distances $\langle r_{ij}\rangle$ are recovered.
The $\boldsymbol{K}$ matrices were obtained using an iterative optimization algorithm described in a previous work \cite{Shi2023NatComm}. The $\boldsymbol{K}$ matrices plotted in fig. S3 show the characteristic banded structure for each polymer model, and we confirmed that each inferred $\boldsymbol{K}$ exactly enforces the input Flory scaling exponent $\nu$ (=1/2, 3/5, or 1/3) (fig. S3).
In this example,  we set the total length of chain to be 1,000, and consider two monomers to be symmetrically located  around the midpoint separating them by a linear genomic distance, $s$. Using $\boldsymbol{K}$, we calculated $G_2(t)$ (Eq. \ref{eq:eq3}) for different models. Fig. \ref{fig:fig2}(a) shows the  $G_2(t)$ for an Rouse chain for different $s$ values. The relaxation times $\tau$, obtained using $G_2(\tau)/G_2(0)=1/e$,  are shown as solid circles in Fig. \ref{fig:fig2}(b). Similarly, $G_2(t)$ for the SAW and FG are calculated (see fig. S4).
Because we assume a purely diagonal mobility tensor with  no hydrodynamic (Zimm) coupling, our model fixes $\theta=1$.  Hence, the scaling relation $\tau_p\sim s^{2\nu+\theta}$ becomes $\tau\sim s^{2\nu+1}$, thus recovering $\tau\sim s^2$ for $\nu=1/2$ (Rouse chain) and $\tau\sim s^{5/3}$ for $\nu=1/3$ (FG).   The full Zimm mobility would give $\theta=\nu$ and thus $\gamma=3\nu$.
Fig. \ref{fig:fig2}(b), showing the dependence of $\tau$ as a function of $s$, establishes that the expected scaling of $s^{2}$, $s^{5/3}$, and $s^{2.2}$ are correctly reproduced for a Rouse chain, the FG, and the self-avoiding chain \cite{Toan2008}, respectively. These calculations show that as long as the 3D polymer structures are known then the relaxation times may be accurately calculated. Needless to say that the dependence of $\tau$ with $s$ for homopolymers may be obtained using well-known scaling arguments without resorting to simulations.  

To further demonstrate that the theory reproduces the correct dynamical properties, we also tested it against the polymer simulations of self-avoiding polymers in both good and poor solvents. We first performed equilibrium Brownian Dynamics simulations of self-avoiding polymers (see Supplementary Materials for details). We then computed the mean distance matrix from the trajectories. Using the mean distance matrices as input, we calculated the connectivity matrix using the maximum entropy principle (Eq. \ref{eq:eq2}). The connectivity matrix could be used to calculate the single-monomer MSD $M_1(t)$ and two-point MSD $M_2(t)$. figs. S5 and S6 show that the theory accurately reproduces the correct dynamics in both good and poor solvents.

\begin{center}
\includegraphics[width=.9\columnwidth]{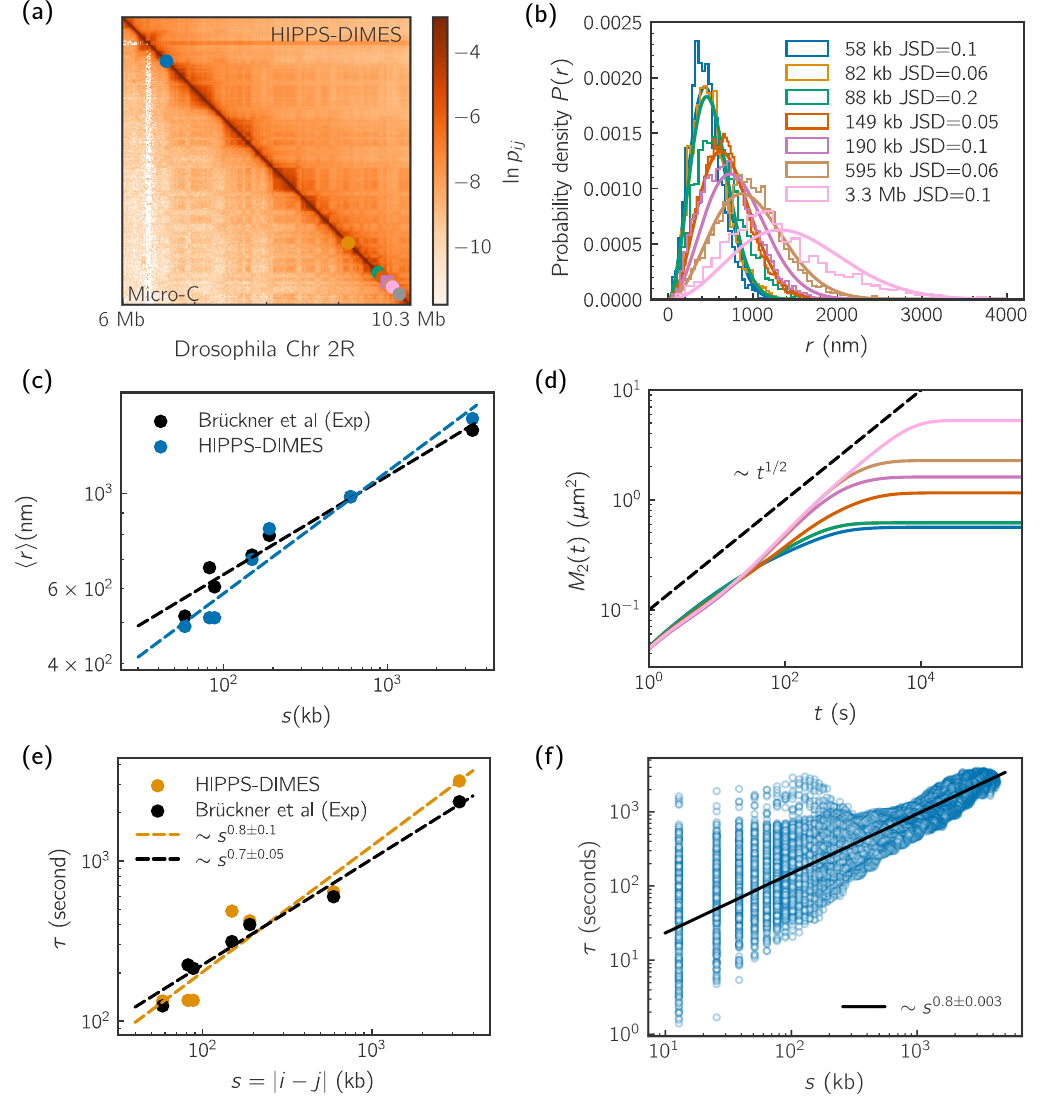}
\captionof{figure}{\label{fig:fig3} \textbf{HIPPS-DIMES accurately captures the static and dynamic properties of \textit{Drosophila} enhancer–promoter pairs}. (a) Comparison between the Micro-C contact map and the predictions using the HIPPS-DIMES method for chromosome 2R in \textit{Drosophila}, marking the promoter (square) and seven different distal enhancers (circles), chosen according to the \textit{eve} promoter-enhancer experimental setup \cite{Bruckner23Science}. (b) Distribution of pairwise distances for the seven promoter-enhancer loci pairs: experimental histograms (bars: data from Ref. \cite{Bruckner23Science}) and HIPPS-DIMES predicted distribution (solid lines). Jensen-Shannon divergence (JSD) values between model predictions and experimental data are reported.
(c) Comparison of the mean spatial distances $\langle r \rangle$ as a function of the genomic distance $s$ between the experimental measurements and the HIPPS-DIMES predictions. (d) Two-point Mean Square Displacement ($M_2(t)$), calculated using $M_2(t) = 2\langle r_{ij}^2 \rangle - 2 G_2(t)$ with $G_2(t)$ from Eq. \ref{eq:eq3}. (e) Comparison of relaxation times $\tau$ as a function of genomic distance $s$ between the experimental observations and HIPPS-DIMES predictions. Black dashed line (Orange dashed line) is the fit to the experimental data (HIPPS-DIMES prediction). (f) Scatter plot for the relaxation time $\tau$ versus genomic separation $s$ for all pairs of loci. Power-law fit is shown in black line.}
\end{center}

\textbf{Application to experiments:} Having established that the theory correctly reproduces the dynamics of a Rouse chain, FG as well as self-avoiding chain, we use it to resolve the conundrum that the equilibrium distances between pairs of loci are incompatible with the observed transcriptional dynamics \cite{Bruckner23Science}. In the Brückner et al. system \cite{Bruckner23Science}, seven engineered enhancer–promoter (E–P) pairs were assayed in \textit{Drosophila} embryos at nuclear cycle 14 (nc14): a minimal eve promoter (MS2‐tagged) at a fixed locus and a synthetic eve enhancer (ParS‐tagged) inserted at defined distances (58 kb, 82 kb, 149 kb, etc.) from the promoter.  To mimic each reporter pair, we use the wild‐type Micro-C contact map for \textit{Drosophila} embryo nc14 cells \cite{IngSimmons2021} and select the locus at the same genomic positions from the endogenous eve promoter as in the experiment, assuming that a single small insertion does not noticeably alter large‐scale contacts. The derivation leading to Eq. \ref{eq:eq3} (see Supplementary Materials for details)  shows that, if all the loci in the chromatin experience the same friction coefficient $\xi$, the dynamics based on the HIPPS-DIMES model is fully determined by the connectivity matrix $\boldsymbol{K}$. The expression in Eq. \ref{eq:eq3} can be numerically computed using the eigenvalues/eigenvectors of $\boldsymbol{K}$ (details are given in the Supplementary Materials). In the HIPPS-DIMES theory,  $\boldsymbol{K}$ for any chromosome may be readily calculated from the static contact map (Hi-C or Micro-C) or the imaging data.  Fig. \ref{fig:fig3}(a)  compares the HIPPS-DIMES prediction for the contact map with the Micro-C data (see fig. S12 for zoom-in view on one Mb-scale comparison). The enhancers and promoter used in the experiment setup \cite{Bruckner23Science} are marked. In addition to reproducing the contact map faithfully, Fig. \ref{fig:fig3}(b) shows that the distributions, $P(r)$, of spatial distance between promoter and seven enhancers are \textit{quantitatively }recovered (Jensen-Shannon Divergence (JSD) between the empirical distributions and model predictions are calculated and shown). Model predictions in Fig. \ref{fig:fig3}(b) were generated by sampling 10,000 independent HIPPS-DIMES conformations and computing all pairwise distances. Note that the 82 kb (orange) and 88 kb (green) traces overlap exactly due to the $12.8\,$kb genomic‐resolution of the reconstructed structures. Fig. \ref{fig:fig3}(c) shows the spatial distances, $\langle r\rangle$, as a function of $s$. These results show that the structural predictions of the HIPPS-DIMES, using the Micro-C contact map as input, are in excellent agreement with both Micro-C and imaging experiments. 

Next, we calculated the two-point mean square displacement $M_2(t)$ using Eq. \ref{eq:eq3} and the relation $M_2(t)=2\langle r_{ij}^2(t)\rangle-2G_2^{ij}(t)$. Use of Eq. \ref{eq:eq3} requires knowledge of the connectivity matrix $\boldsymbol{K}$, which is the byproduct of the determination of the 3D chromatin coordinates (Eq. \ref{eq:eq2}).  Fig. \ref{fig:fig3}(d) shows the time dependence of $M_2(t)$ predicted by our theory for the pairs of E-P distances. At long times, $M_2(t)$ saturates, approaching the different equilibrium values that depend on the given pair. The rate of approach depends on the specific enhancer and promoter pair.

We then calculated the relaxation time $\tau$. Because the absolute value of the friction coefficient is unknown, we tuned it to achieve the best agreement between the theoretical prediction and experimental value for $\tau$.  The fit parameter yields the unit length  $l_0=147\mathrm{nm}$ and unit time is $\tau_0\approx 3.1\mathrm{s}$. Using $\tau_0=3.1$s and $l_0=140$nm, one finds $\xi = k_{\mathrm{B}}T\,\tau_0/l_0^2\approx6.5\times10^{-7}\,\mathrm{N\,s/m}$. Using Stokes' law with $r=l_0/2$, we estimate the viscosity of the environment to be $\eta=\xi/(6\pi r)\approx0.5\,$Pa·s. We used these values to calculate the theoretical predictions for  $\tau$ versus genomic distance $s$ to compare with experiments.  The theoretical prediction for the scaling exponent, $\gamma$ in $\tau \sim s^{\gamma}$,  is $\approx 0.8$ and for experimental data is $\approx 0.7$ (Fig. \ref{fig:fig3}(e)). It is important to note that while $l_0$ and $\tau_0$ are adjustable parameters used to calculate the absolute values of $\tau$, they do not affect the scaling exponent $\gamma$.  We consider the agreement for $\tau$, and especially $\gamma$, between experiment and theory to be striking because the only information that is used in the calculation is the Micro-C static contact map.

\textbf{Randomly Shuffled Sequence:} We then wondered if the introduction of randomness in the Micro-C contact map would lead to a discrepancy between theory and experiment. To this end, we randomly shuffled the pairwise distances in the distance map but retained the first off-diagonal elements. In this way, the polymeric nature of the structure is preserved, but the specific WT (wild type) pattern in the Micro-C contact map is destroyed. We then applied HIPPS-DIMES on the shuffled distance map to obtain the connectivity matrix. Comparison of the contact maps between the WT and the randomly shuffled sequence shows (fig. S7(a)) that the pattern in the WT contact map is fully disrupted. fig. S7(b) shows the mean pairwise distances as a function of the genomic distance. For $s<10^2$ bps, the mean distance grows roughly as a power law with an exponent of 0.6, demonstrating the polymeric nature is preserved in the random shuffled system. At $s>10^2$ bps, the mean distances reach the plateau, which is a result of random shuffling. The relaxation time $\tau$ for the shuffled sequence is insensitive to the genomic distance (fig. S7(c)), which is consistent with the saturation of mean distances. The purpose of this calculation is to show that the scaling of $\tau$ as a function of $s$ cannot be captured in random heteropolymer. The chromatin sequence, reflecting the patterns of activity depicted in histone modifications, and the associated 3D structures should be accounted for precisely.

\begin{figure}
\includegraphics[width=\columnwidth]{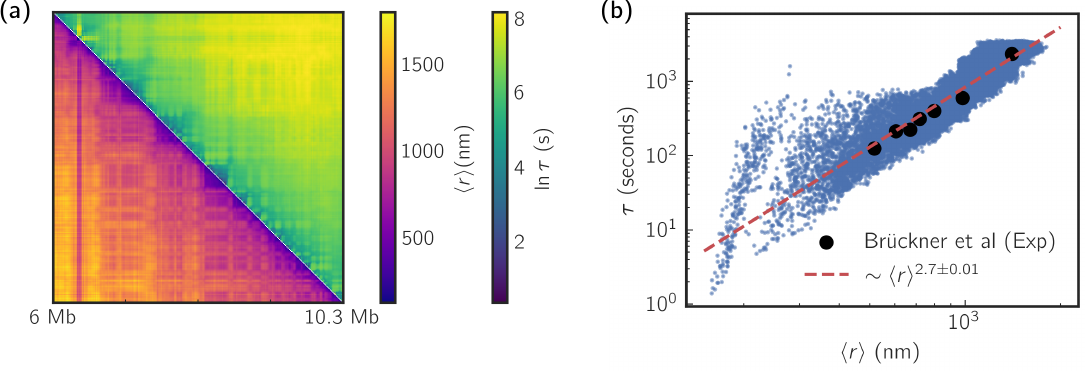}
\caption{\label{fig:fig4} \textbf{Locus‐specific relaxation times scale with spatial separation of chromatin loci in \textit{Drosophila} embryo cells}. (a) Heatmap comparison of the mean spatial distances  (lower triangle) and relaxation time $\tau$ (upper triangle) across all locus pairs. (b) Log–log scatter plot of the relaxation time $\tau$ versus mean spatial distance $\langle r\rangle$. Black circles are experimental measurements; blue squares are HIPPS-DIMES predictions. The red dashed line is a power-law fit  $\tau \sim \langle r\rangle^{2.7}$ highlighting the non-Rouse scaling. }
\end{figure}

\textbf{Loci-dependent relaxation times:} Given the remarkable success of our theoretical approach in quantitatively explaining the experimental findings, we calculated all the pairwise relaxation time $\tau_{ij}$ where $i,j$ are the loci pair indices.  The value of $\tau_{ij}$ depends on both $i$ and $j$, and not merely on the genomic distance $s=|i-j|$ as in the case of a homopolymer. On an average, the relaxation time correlates with both $s$ and the mean spatial distance $\langle r_{ij}\rangle$ in a non-trivial manner. Fig. \ref{fig:fig4}(a) shows the mean spatial distance map and the relaxation time map, clearly establishing the correlation between the two quantities (see fig. S13 for zoom-in view of the one Mb scale comparison). Fig. \ref{fig:fig4}(b) shows the scatter plot of $\langle r_{ij}\rangle$ versus $\tau_{ij}$. The results show that, on an average, they are related as $\tau_{ij}\sim \langle r_{ij}\rangle^{2.7}$, which differs substantially from the prediction for the Rouse  and the FG models.  In particular,  the scaling exponent $\sim 2.7$ is substantially smaller than the Rouse model prediction ($=4$) and the value for FG ($=5$).

We then wondered whether the observed scaling can be deduced by considering an effective homopolymer model, in which the mean distance matrix is calculated as the average of the wild-type (WT) distance map over fixed genomic distances. The calculation is intended to assess if a modified scaling relation could be used with the mean distance that is calculated from the contact maps. We computed $r(s)$ by averaging $\langle r_{ij}\rangle$ over $s = |j - i|$, and then applied HIPPS-DIMES to obtain the relaxation time. fig. S8 shows the calculated average distance map, demonstrating that $r(s) \sim s^{1/4}$ for $s > 10^2\ \mathrm{kb}$. figs. S8(c) and 8(d) show that the relaxation time scales with genomic distance as $\tau \sim s^{1.1}$ and with mean pairwise distance as $\tau \sim r^{4}$. Both of these scaling relations are different from the results obtained by considering the full WT contact map. This further demonstrates that arguments in standard polymer physics for homopolymer do not hold for the chromatin, thus  underscoring the importance of considering the complete information embedded in the WT contact map. Together, these results show that the relaxation process between a pair of chromatin loci is much faster than predicted by standard polymer models, which provides a structural basis for interpreting the key experimental finding~\cite{Bruckner23Science}. Importantly, the rapid dynamics between distal loci can be explained by taking into account the actual 3D coordinates of the chromosomes.

Finally, we examined whether enhancer-promoter (E-P) relaxation is faster than that of non-E-P pairs at the same genomic distances. For each of the six reporter separations (at 58 kb, 82 kb, 149 kb, etc.), we calculated the distribution of relaxation times for all loci pairs separated by the same genomic separation. The results, shown in fig. S9, reveal distinct trends depending on the on the genomic distance. For very short genomic distances (enhancer-promoter pairs 1 and 2, with separations of 58 kb and 82 kb), the mean relaxation times (133 s and 134 s, respectively) are substantially shorter than the mean relaxation times of all pairs at the same genomic distance (207 s and 240 s). In contrast, for larger genomic distances (all other enhancer-promoter pairs), the mean relaxation times are consistently longer than the corresponding mean values of all pairs with the same separation. Because  these specific pairs are only available in the current experimental measurements, we cannot generalize further to other E–P combinations.

\begin{figure}
\includegraphics[width=\columnwidth]{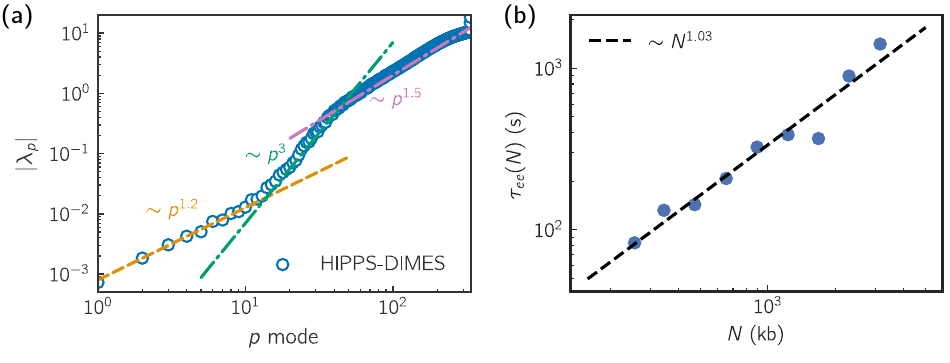}
\caption{\label{fig:fig5} \textbf{Eigenvalue spectrum and chain‐length dependence of the end-to-end relaxation times in \textit{Drosophila} embryo cells}. (a) Scaling of the eigvenvalues, $|\lambda_p|$, of the connectivity matrix as a function of mode index $p$, revealing three distinct scaling regimes. (b) End-to-end relaxation time $\tau_{ee}$ as a function of chain length $N$. The dashed line shows a power-law fit.}
\end{figure}

\textbf{Plausible mechanism for rapid relaxation dynamics in chromatin:} To explore the underlying mechanism for the rapid relaxation times between chromatin loci, we calculated the spectrum of eigenvalues, $\lambda_p$, associated  with, $\boldsymbol{K}$, the connectivity matrix. Interestingly, Fig. \ref{fig:fig5}(a) shows that the scaling  of $|\lambda_p|$ with $p$ has a complex structure. There are three distinct regimes in the variation  of $\lambda_p$ with the normal mode $p$.  For $p\leq 10$,  we find that $|\lambda_p| \approx p^{1.2}$.  In the second regime, $10\leq p \leq 50$ the eigenvalues increase as $|\lambda_p| \sim p^3$. Finally,  $|\lambda_p| \sim p^{1.5}$ for $p \geq 50$. The complicated spectrum for chromatin should be contrasted with the Rouse model for which  $|\lambda_p|\sim p^2$ where the inverse of $|\lambda_p|$ maybe interpreted as the relaxation time of $N/p$ segments of the chain. Notably, the smaller scaling exponents in the small $p$ regime (Fig. \ref{fig:fig5}(a)) compared to the $p^2$ scaling of the Rouse model 
supports the finding that chromatin relaxes more rapidly, consistent with the results for relaxation time $\tau$. In general, for a homopolymer, we expect that `$|\lambda_p|\sim p^x$,  which implies that the end-to-end relaxation times ($\tau_{ee}$) scale with respect to the chain length with the same exponent, $\tau_{ee}\sim N^x$. For instance, for the Rouse model, $|\lambda_p|\sim p^2$ and $\tau_{ee}\sim N^2$. If we assume that a similar power law relationship holds between $\tau_{ee}$ and $p$ in chromosomes, then expect that $\tau_{ee}\sim N^{1.2}$ if $s \gtrapprox$ 400 kb. To test this prediction, we calculated the end-to-end-distance  relaxation times in chromosomes with  different lengths. The HIPPS-DIMES-based calculation shows that the end-to-end relaxation time roughly scales linearly with  $N$  (Fig. \ref{fig:fig5}(b)) as $\tau_{ee}$ varies by over five orders of magnitude. We also computed the eigenvalues for the randomly shuffled system. fig. S10(a) shows that $|\lambda_p|$ becomes independent of $p$ for $p\leq 40$, which is is consistent with the findings that $\tau$ is insensitive to the genomic distance in the shuffled system. Together, these results suggest that the dynamics of chromosomes are dependent on the sequence and the length scale, which is reflected in the observation that the $|\lambda_p|$  exhibits three distinct scaling regimes at different $p$ (different length scales).

\begin{figure}
\includegraphics[width=\columnwidth]{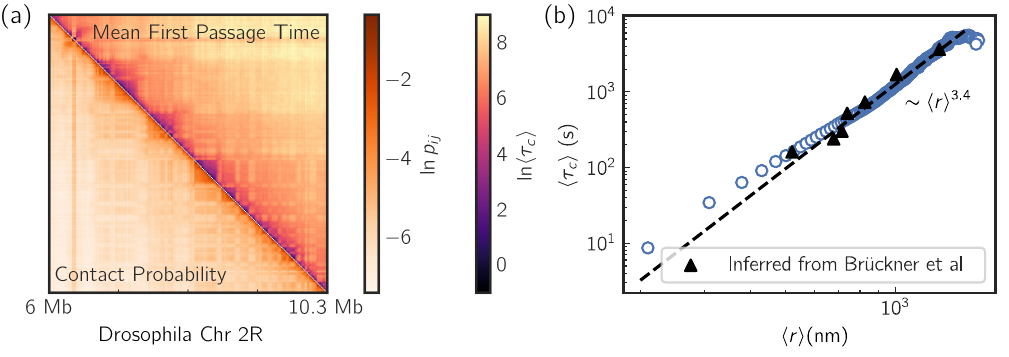}
\caption{\label{fig:fig6} \textbf{First‐passage times for contact between chromatin loci in \textit{Drosophila} embryo cells.} (a) Comparison between heatmap for the mean first-passage time of contact between chromatin loci (upper triangle), $\langle \tau_c\rangle$, and the contact map (lower triangle). The contact is defined with a distance threshold of $r_c=147\mathrm{nm}$. (b) $\langle \tau_c\rangle$ scales as the mean distance $\langle r\rangle$ as a power-law. The dashed line is a guide for the eye. Triangle symbols are the data computed using experimental trajectory data \cite{Bruckner23Science} (See Supplementary Materials).}
\end{figure}

\textbf{First-passage time of contact between distal loci}: A functionally relevant biophysical property related to the two-point relaxation time is the first-passage time of contact between a pair of chromatin loci. A simpler, well-studied, and instructive version of this problem in polymer physics is the cyclization process, which concerns the first-passage time for two ends of a polymer chain to meet \cite{Wilemski1974, Szabo1980,Gurin2012, Toan2008}. Let us denote the first-passage time of contact as $\tau_c$, which is determined by the search process by which two loci meet. It can be shown that $\tau_c$ is directly connected to the two-point relaxation and is governed by the relaxation dynamics between the loci \cite{Toan2008, Gurin2012}, assuming that the threshold for establishing contact is not small. Using a contact threshold of $r_c=l_0=147\mathrm{nm}$, we estimated the mean FPT, $\langle \tau_c\rangle$, for all pairs of the chromatin loci (see Supplementary Materials for method to estimate mean FPT). Fig. \ref{fig:fig6}(a) shows that the domain along the diagonal in the $\tau_c$ map visually matches the contact domains in the contact map. We then calculated  $\tau_c$ by averaging over the spatial distance $r$ with fixed genomic distance $s$. Fig. \ref{fig:fig6}(b) shows that, on an average, the mean $\tau_c$ of contact between two chromatin loci scales as $\langle r \rangle^{3.4}$.

We also compared our theoretical predictions with results computed using experimental loci trajectory data. The method for calculating the mean FPT from experimental trajectories is described in the Supplementary Materials. As shown in Fig. \ref{fig:fig6}(b), the experimental results (represented by triangle symbols) align closely with our predictions, demonstrating excellent quantitative agreement between the experiment and the theory. 

Interestingly, the calculated scaling exponent of 3.4 is close to 3, which is the theoretical prediction by Szabo, Schulten, and Schulten (SSS) \cite{Szabo1980} who derived the first-passage time of contact under the assumption that two-point diffusion is governed by dynamics in a potential of mean force, which can be calculated analytically for the Rouse model. The potential of mean force is determined from the distribution of distances between two loci which further leads to \cite{Szabo1980, Toan2008, Kwon2017prl},

\begin{equation}\label{eq:eq4}
    \tau_{c, \mathrm{SSS}}=\frac{1}{D}\int_{r_c}^{L}\mathrm{d}x\frac{1}{p(x)}\bigg(\int_x^{L}\mathrm{d}y p(y)\bigg)^2
\end{equation}

\noindent where $p(x)$ is the equilibrium distribution of distances $x$ between the two loci with the mean distance $\langle r\rangle$, $r_c$ is the threshold distance distance for contact, and $D$ is the effective diffusion constant. It can be shown using Eq. \ref{eq:eq4} that $\tau_{c,\mathrm{SSS}}\sim D^{-1} r_c^{-1} \langle r\rangle^3$. This result is consistent with the visual similarity between the contact map and the contact time map shown in Fig. \ref{fig:fig6}(a), as the contact probability scales with the mean distance with an exponent of 3 (in the HIPPS-DIMES model).  It is intriguing that although the SSS theory fails to predict the correct scaling in the Rouse model, it provides a better description of chromatin dynamics.

The value of $\tau_c$ is important in describing the dynamics of enhancer-promoter (EP) communication and potentially in understanding EP-regulated gene expression. If we assume that gene expression is initiated by the formation of contact between promoter and enhancer, then the transcription rate can be expressed as $k=1/(\tau_{\mathrm{d}} + \tau_c)$, where $\tau_{\mathrm{d}}$ is the time required for the downstream processes that ensue after the establishment of contact. By expressing $\tau_c$ as a function of the contact probability $p_c$, with $\tau_c \sim \tau_0 p_c^{-\theta}$, we obtain $\hat{k}=k/k_{\mathrm{max}} = 1/(1 + (\tau_0/\tau_d) p_c^{-\theta})$. This equation can be considered as the dynamic analog of the Hill equation with the cooperativity parameter $\theta$.  Such an equation has been used to model the mean mRNA number as a function of the contact probability between promoter and enhancer \cite{Zuin2022}.

\begin{figure}
\includegraphics[width=\columnwidth]{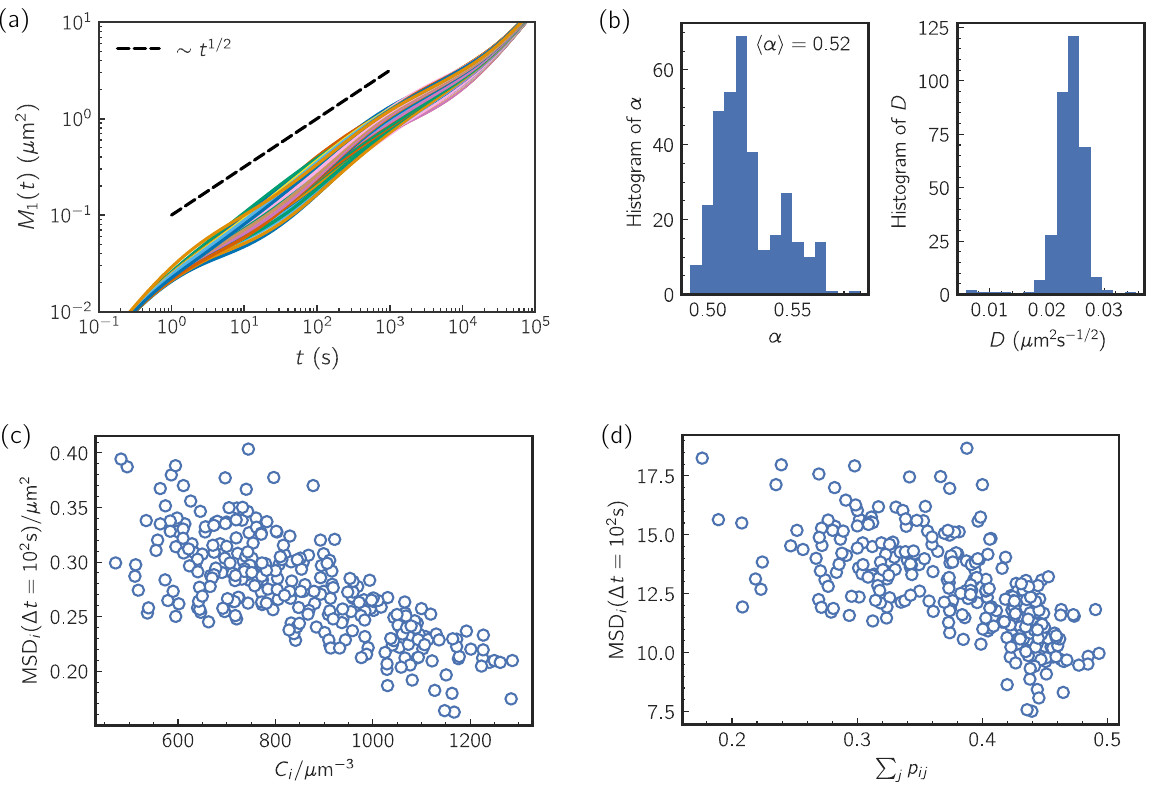}
\caption{\label{fig:fig7} \textbf{Single-locus dynamics and their dependence on chromatin network centrality in \textit{Drosophila} embryo cells.} (a) Single-locus mean square displacement $M_1(t)$. Each curve corresponds to an individual locus. (b) Histogram of the fitted diffusion exponent $\alpha$ and diffusion coefficients $D$. (c) Locus-specific diffusivity, defined as $M_1(t=10^2\ \mathrm{s})$, versus closeness centrality $C_i = \sum_{j\neq i} \langle r_{ij}\rangle^{-m}$ with $m=3$. (d) Scatter plot of locus‐specific diffusivity, defined as $M_1(t)$ at $t=10^2$ s, versus total contact connectivity $\sum_{j\neq i}p_{ij}$ as a function of the sum of contact probabilities for each locus $i$.}
\end{figure}

\textbf{Single-Locus Dynamics and Centrality Measure:} Having demonstrated that our theory quantitatively reproduces the experimental data for the two-point dynamics in chromatin loci,  we explore the predictions of single-locus dynamics. The single-locus mean square displacement, $M_1(t)$, is computed using Eq. 14 in the Supplementary Materials. The prediction for $M_1(t)$ is shown in Fig. \ref{fig:fig7}(a), where each line represents a single chromatin locus. Between the time scale of $1 \mathrm{\ s} < t < 10^5\mathrm{\ s}$, $M_1(t)$ scales approximately with an exponent of $\sim 0.5$, which is close to the prediction of the Rouse model. Next, we calculated the diffusion exponent $\alpha$ and diffusion constant $D$ for each locus by fitting $M_1(t)$ in the time range $1 \mathrm{s} < t < 10^5\mathrm{s}$ using $M_1(t) = D t^{\alpha}$. The histogram of $\alpha$ and $D$ reported in Fig. \ref{fig:fig7}(b), yields the loci averages of $\langle \alpha\rangle = 0.52$ and $\langle D\rangle = 0.024\ \mu\mathrm{m^2 s^{-1/2}}$. Fig. \ref{fig:fig7}(b) shows broad distributions of both the exponent $\alpha$ and the effective diffusion coefficient, indicating that single‐locus diffusion is heterogeneous.

We then investigated the factors that determine the variance in single-locus diffusion. We hypothesize that chromatin loci should generally diffuse more slowly if they have higher local density. Inspired by concepts in graph theory~\cite{Newman12NatPhys}, we defined the closeness centrality measure based on the mean pairwise distances. Let us define the centrality of a single locus as $C_i$, where $i$ is the locus index, as, $C_i = \sum_{j\neq i} \langle r_{ij}\rangle^{-m}$, where $\langle r_{ij}\rangle$ is the mean pairwise distance between the $i^{\mathrm{th}}$ and $j^{\mathrm{th}}$ loci, and $m > 0$ is an adjustable parameter. The centrality of a locus is higher when it is in proximity to other loci. Using $m=3$, we plotted $C_i$  against $M_1(t)$ at $t=10^2\mathrm{s}$ (Fig. \ref{fig:fig7}(c)). The results show a negative correlation between the diffusivity of loci and the centrality measure. We recognize that the inverse of the pairwise distance $r_{ij}$ is correlated with the contact probability $p_{ij}$. We then inspect the correlation between diffusivity $M_1(t=10^2\ \mathrm{s})$ and $\sum_{j\neq i}p_{ij}$ and find that these two quantities are indeed anticorrelated (Fig. \ref{fig:fig7}(d)).

\begin{figure}[H]
\includegraphics[width=\columnwidth]{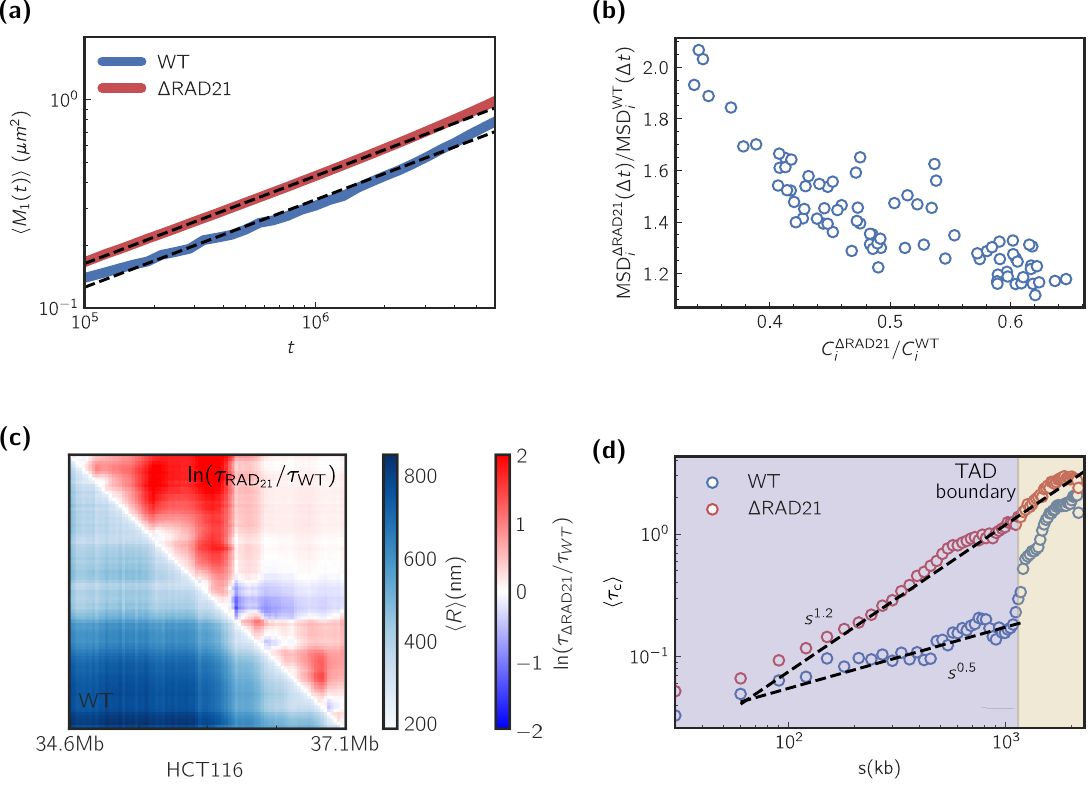}
\caption{\label{fig:fig8} \textbf{Cohesin depletion accelerates chromatin single-locus mobility while either prolonging or accelerating inter-locus relaxation in human HCT116 cells.} (a) Loci-average single-locus mean square displacement $\langle M_1(t)\rangle $ for wild-type (WT) and cohesin-depleted ($\Delta \mathrm{RAD21}$) chromosomes. $\langle M_1(t)\rangle$ for each case is calculated by averaging the single-locus $M_1(t)$ over all the chromatin loci.  Dashed lines are fits to the data within the time window shown in the figure, with diffusion exponent of $\alpha=0.41$ for WT and $\alpha=0.43$ for $\Delta$RAD21. (b) Relative change in diffusivity after cohesin deletion vs. relative change in centrality. (c) Lower triangle: mean distance map for WT chromosomes. Upper triangle: relative change in relaxation time after cohesin deletion. (d) Mean first‐passage time to contact, $\langle \tau_c\rangle$, plotted against genomic separation $s$ for WT (blue) and $\Delta$RAD21 (red) chromosomes. Contacts are defined when loci approach within 200 nm. Purple and yellow shaded regions denote contacts occurring within versus outside TADs. Dashed lines are power‐law guide to the eye.}
\end{figure}

\textbf{Effect of Cohesin Deletion:} The generality of our theory allows us to predict the consequences of deleting cohesin on the chromatin loci dynamics. The ATP consuming motor, cohesin, extrudes loops \cite{Ganji2018science, Dekker2023science} in interphase chromosomes, which results in the formation of Topologically Associating Domains (TADs) revealed in the Hi-C contact map \cite{Sanborn2015pnas, Fudenberg2016}. We took advantage of the imaging data \cite{Bintu18Science} and applied the HIPPS-DIMES method to the experimentally measured mean distance map of human Chromosome 21 for both the wild-type (WT) and cohesin-depleted ($\Delta \mathrm{RAD21}$) HCT116 cell lines. After determining the connectivity matrix $\boldsymbol{K}$ by using the measured distance map as constraints, we calculate $M_1(t)$ for each chromatin locus. We then computed the locus-averaged mean square displacement using $\langle M_1(t)\rangle=(1/N)\sum_i M_1^{i}(t)$, where $N$ is the total number of loci and $M_1^{i}(t)$ is the mean square displacement for locus $i$. Fig. \ref{fig:fig8}(a) shows that chromatin loci in cohesin-depleted ($\Delta \mathrm{RAD21}$) cells have higher diffusivity (diffusion coefficients) than in the wild-type (WT) cells. The increase in diffusivity ranges between 20\% to 40\% on the time scales of $10^5 < t < 10^7$. We cannot estimate the absolute  value for the time scale because of lack of reference experimental data to benchmark the theory. Therefore, time is reported in reduced units. We then calculated the relative change in diffusivity as a function of the relative change in centrality. Fig. \ref{fig:fig8}(b) shows that the centrality of chromatin loci decreases after cohesin deletion. Loci exhibiting a greater reduction in centrality typically show a larger percentage increase in diffusivity. Since the diffusivity of loci is anticorrelated with their centrality, these results show that cohesin-mediated loop extrusion constrains loci dynamics. As a consequence, its deletion leads to increased single-locus diffusivity---a prediction that is in quantitative agreement with experiments \cite{Mach2022, Gabriele2022}. 

Next, we investigate the two-point loci dynamics by calculating the relaxation time $\tau$. Fig. \ref{fig:fig8}(c) shows a heatmap of the relative change in relaxation time between $\Delta$RAD21 and WT cells, $\ln (\tau_{\Delta \mathrm{RAD21}} / \tau_{\mathrm{WT}})$. Fig. \ref{fig:fig8}(c) shows that although chromatin single-locus dynamics are accelerated after cohesin deletion, the change in two-point relaxation time $\tau$ is not uniform but is locus-dependent. For loci located within the TADs, the relaxation time increases after cohesin deletion because the distances between the loci increase. In contrast, loci located at the boundaries of TADs, the relaxation time decreases after cohesin deletion. These predictions are amenable to experimental tests. Mechanistically, cohesin depletion disrupts loop‐extrusion–mediated insulation at TAD borders, allowing boundary‐spanning loci to come into proximity and relax more quickly, while loci within TADs lose compaction and relax more slowly. Finally, we computed the mean first‐passage time to contact, $\langle\tau_c\rangle$, as a function of genomic separation $s$ (Fig. \ref{fig:fig8}(d)). In WT cells, loci within the same TAD obey $\langle\tau_c\rangle\sim s^{0.5}$, but pairs crossing TAD boundaries exhibit a sharp jump to much larger $\tau_c$. In contrast, in cohesin-depleted $\Delta$RAD21 cells $\langle\tau_c\rangle$ scales as $s^{1.2}$, with virtually no distinction between within‐TAD and outside‐TAD contacts — consistent with the loss of loop‐extrusion–mediated insulation. This suggests that loop extrusion greatly reduces the time required for loci to contact within TADs.

We also computed the mode spectrum---characterized by the eigenvalues $\lambda_p$---for both wild-type (WT) and cohesin-depleted cells.  fig. S10(b) shows that the cohesin-depleted system exhibits a distinct gap between the $p=1$ and $p=2$ modes, suggesting a separation of relaxation time scales between the entire region and a subset of the region. This gap likely arises from the disruption of two loop extrusion domains present in the WT. Upon cohesin depletion, these domains disappear, and the separation between the $p=1$ and $p=2$ modes vanishes. Moreover, we observed a qualitative change in the scaling behavior of $|\lambda_p|$ with respect to $p$: in WT cells, $|\lambda_p| \propto p^{1.2}$, whereas in the cohesin-depleted cells, $|\lambda_p| \propto p^{1.5}$. The steeper scaling observed in the cohesin-depleted system indicates that small-scale fluctuations dissipate more rapidly, suggesting a loss of coordinated motion across larger chromatin domains.

\newpage

\section{Discussion}
In this study, we developed a theoretical framework to predict chromatin dynamics from ensemble-averaged static contact maps to make a precise connection between three-dimensional structure and dynamics. By employing the HIPPS-DIMES methods, we reconstructed the three-dimensional structures of chromatin using the experimental contact maps and derived the connectivity matrix $\boldsymbol{K}$, which encapsulates effective pairwise interactions between chromatin loci. Interpreting this matrix within the context of polymer dynamics allowed us to compute the dynamical correlation functions and predict chromatin dynamics using a generalized Rouse model framework.

Our theory, with no locus-specific fitting parameter and containing a single adjustable parameter — the effective friction coefficient that  sets the overall time scale —  accurately reproduces the experimental observations of loci dynamics in \textit{Drosophila} embryo cells,  thus resolving the apparent discordance between static chromatin structures and dynamic behaviors highlighted in recent studies~\cite{Bruckner23Science}. Strikingly,  the two-point relaxation times between chromatin loci scale with genomic separation is in excellent agreement with experiments demonstrating that the unexpected rapid relaxation dynamics maybe a consequence of effective long-range interactions, which could be mediated by factors like transcription factors and cohesin. By analyzing the eigenvalues of the connectivity matrix, we uncovered that the rapid chromatin dynamics exhibit complex, length-scale-dependent behavior, which may be connected to the hierarchical structural organization of chromosomes. This finding suggests that the dynamics of chromosomes cannot be captured by homopolymer models but requires knowledge of the intricate network of interactions in chromatin. A concise analytical connection linking eigenvalue spectrum to the dynamic scaling exponents remains an important avenue for future theoretical work. We also used our model to predict the mean first-passage times for contact between chromatin loci and found quantitative agreement with results calculated using the experimentally measured loci trajectories~\cite{Bruckner23Science}. This calculation further supports the notion that chromatin dynamics are intrinsically linked to the precise static three-dimensional structure, which likely play a crucial role in processes such as promoter-enhancer communication.

Additionally, our theory predicts that the heterogeneous single-locus diffusion behavior is dependent on local chromatin density. We found that loci with higher contact probabilities with other loci tend to exhibit slower diffusion, highlighting the influence of the interaction landscape on chromatin mobility. Our exploration of the effects of cohesin deletion revealed that chromatin loci in cohesin-depleted cells exhibit higher diffusivity and the changes in the two-point relaxation times are locus-dependent. This observation underscores the role of cohesin in regulating chromatin dynamics.

Although our main analyses focus on \textit{Drosophila} chromosome 2R, we have also applied HIPPS-DIMES to human HCT116 cells to study cohesin depletion, demonstrating its direct applicability to mammalian chromosomes. In principle, the framework is agnostic to species or cell line. To underscore the generality of HIPPS‐DIMES, we have also applied the theory to human GM12878 Hi‐C data (Supplementary Materials and fig. S14) and to mESC Micro‐C data at the Fbn2 locus (fig. S15), with comparable success in reproducing both single‐locus and two‐point dynamics. However, its current implementation treats each chromosome as a separate, contiguous polymer. Modeling on the whole-genome level would require either independent per-chromosome analyses or an extension to capture inter-chromosomal contacts.

It is important to recognize that HIPPS-DIMES relies on experimental data (Hi-C/Micro-C or imaging) as input, so any noise or error in these data will propagate into our predictions. Sources of error include uneven mappability and fragment-level biases in Hi-C/Micro-C, fixation artifacts introduced by Hi-C/Micro-C and many chromosome-tracing protocols (e.g., formamide-based FISH) \cite{IrgenGioro2022}, and localization uncertainty in imaging data \cite{Brando2021, Yang2024, Bohrer2025}. To assess the impact of such errors, we verified that our dynamical predictions remain robust to moderate Hi-C/Micro-C noise and to realistic imaging localization errors ($\leq$ 50 nm); see Supplementary Materials and figs. S16–17. In future work, it would be valuable to extend our framework to explicitly model localization uncertainty or to apply HIPPS-DIMES to data generated by non-denaturing methods such as RASER-FISH \cite{Brown2022, Beckwith2021, Beckwith2025}.

In summary, by moving beyond homopolymer models like the Rouse and fractal globule models, which predict relaxation exponents that are inconsistent with experiments and fail to capture the rapid locus‐dependent dynamics, we show that the precise, heterogeneous 3D structure inferred by HIPPS‐DIMES dictates the observed chromatin dynamic behavior. The proposed theoretical framework, which could be applied to other systems, resolves the conundrum raised in the experiments~\cite{Bruckner23Science}. Importantly, we have shown that measurements of the contact map using Hi-C/Micro-C or the distance map using the imaging method are sufficient to calculate the loci-specific chromatin dynamics. Because HIPPS‐DIMES predicts time‐resolved trajectories for chromatin loci, the predictions can be directly validated using live‐cell tracking methods—such as CRISPR/dCas9 tagging, operator‐repeat arrays, or MS2/MCP reporters—by comparing properties such as mean‐square displacements, relaxation times, and first‐passage statistics. By establishing a direct and quantitative link between chromatin structure and dynamics, our general theoretical framework opens avenues for exploring chromatin dynamics in various biological contexts.

\AtEndDocument{%
\nocite{Gillespie1996, Thompson2022, Weeks1971, Kremer1990, klein2006survival, Harris2023, Jusuf2025}
}

\bibliography{main}

\section*{Acknowledgement}
We are grateful to Mauro L. Mugnai for useful comments and discussions.

\textbf{Funding}: This work was supported by a grant from the National Science Foundation (CHE 2320256, recipient D.T.) and the Welch Foundation through the Collie-Welch Chair (F-0019, recipient D.T.).

\textbf{Author contributions}: G.S. and D.T. designed the research. G.S. performed the research. G.S., S.S. and D.T. analyzed the data. G.S., S.S., and D.T. wrote the manuscript.

\textbf{Competing Interests}: The authors declare no competing interests. 

\textbf{Data and Materials Availability}: We used the following publicly available data: Micro-C data for \textit{Drosophila} embryos are available from ArrayExpress (E-MTAB-9784; \url{https://www.ebi.ac.uk/biostudies/arrayexpress/studies/E-MTAB-9784}), multiplexed FISH imaging data for HCT116 cell line from GitHub (\url{https://github.com/BogdanBintu/ChromatinImaging}), Hi-C data for human GM12878 cells from ENCODE (ENCFF065LSP; \url{https://www.encodeproject.org/files/ENCFF065LSP/}), and Micro-C data for mESCs from GEO (GSE286495). The live-cell imaging data from Bruckner et al. \cite{Bruckner23Science} is publicly available at Zenodo repository \cite{david_bruckner_2023_7616203} at \url{https://zenodo.org/records/7616203}.
To process Hi‐C/Micro‐C data in \textit{.cool} and \textit{.hic} formats, we used the \textit{Cooler} Python package \cite{Abdennur2019} at \url{https://github.com/open2c/cooler} and the \textit{hic‐straw} Python package \cite{Durand2016} at \url{https://github.com/igvteam/hic-straw}, respectively. The code for the model presented in this work and its detailed user instructions can be accessed at the Zenodo repository \cite{guang_shi_2025_15611946} at \url{https://zenodo.org/records/15611946} or at GitHub repository for the latest version: \url{https://github.com/anyuzx/HIPPS-DIMES}. The data analysis is performed using Python 3.12 in Jupyter Lab. The Python packages used in analyzing data and visualization are Scipy, Numpy, Pandas, and Matplotlib. All data needed to evaluate the conclusions in the paper are present in the paper and the Supplementary Materials.

\section*{Supplementary Materials}
\begin{itemize}
    \item Supplementary Text
    \item Figs. S1 to S17
\end{itemize}

\newpage

\end{document}